\documentclass[twocolumn,floatfix, nofootinbib,prd,preprintnumbers,showpacs,showkeys,superscriptaddress,longbibliography]{revtex4-1}

\usepackage[mathscr]{euscript}
\usepackage{amsmath}
\usepackage{graphicx}
\usepackage{dcolumn}
\usepackage{bm}
\usepackage{epsfig}
\usepackage{amssymb,latexsym,mathrsfs}
\usepackage{graphicx}
\usepackage{color}
\usepackage{hyperref}
\usepackage{float}
\usepackage{diagbox}

\usepackage{tikz}

\usepackage{amsmath}
\usepackage{physics}
\usepackage{csquotes} 
\newcommand*\diff{\mathop{}\!\mathrm{d}}

\hypersetup{
    colorlinks=true,
    linkcolor=red,
    citecolor=blue,
}

\usepackage{amsmath}
\usepackage{amssymb}
\usepackage{subfigure}
\usepackage{hyperref}
\usepackage{url}
\usepackage{xcolor}
\usepackage{color}
\definecolor{amaranth}{rgb}{0.9, 0.17, 0.31}
\definecolor{purple(munsell)}{rgb}{0.62, 0.0, 0.77}
\definecolor{americanrose}{rgb}{1.0, 0.01, 0.24}
\definecolor{palatinateblue}{rgb}{0.15, 0.23, 0.89}
\definecolor{royalblue(web)}{rgb}{0.25, 0.41, 0.88}
\definecolor{hanpurple}{rgb}{0.32, 0.09, 0.98}
\definecolor{beaublue}{rgb}{0.74, 0.83, 0.9}
\definecolor{carminered}{rgb}{1.0, 0.0, 0.22}
\definecolor{brightpink}{rgb}{1.0, 0.0, 0.5}
\definecolor{vividviolet}{rgb}{0.62, 0.0, 1.0}
\hypersetup{ linktoc=all,
    colorlinks, linkcolor={palatinateblue},
    citecolor={brightpink}, urlcolor={amaranth}}

\newcommand{\be}{\begin{equation}}
\newcommand{\ee}{\end{equation}}
\newcommand{\bs}{\begin{split}} 
\newcommand{\bea}{\begin{eqnarray}}
\newcommand{\eea}{\end{eqnarray}}

\newcommand{\ievlev}[1]{\textcolor{blue}{[{\bf EI}: #1]}}

\newcommand{\bes}{\begin{subequations}}
\newcommand{\ees}{\end{subequations}}

\newcommand{\bo}{\raise-1mm\hbox{\Large$\Box$}}

\begin{document}

\title{Non-thermal photons and a Fermi-Dirac spectral distribution}
\author{Evgenii Ievlev}
\email{evgenii.ievlev@nu.edu.kz}
\altaffiliation[On leave of absence from: ]{National Research Center “Kurchatov Institute”, Petersburg Nuclear Physics
Institute, St.\;Petersburg 188300, Russia}
\affiliation{Physics Department \& Energetic Cosmos Laboratory, Nazarbayev University,\\
Astana 010000, Qazaqstan}
\affiliation{Almaty University of Power Engineering and Telecommunications,\\ 
Almaty 050013, Qazaqstan}
\author{Michael R.R. Good}
\email{michael.good@nu.edu.kz}
\affiliation{Physics Department \& Energetic Cosmos Laboratory, Nazarbayev University,\\
Astana 010000, Qazaqstan}
\affiliation{Leung Center for Cosmology and Particle Astrophysics,
National Taiwan University,\\ Taipei 10617, Taiwan}

\begin{abstract} 
Although non-intuitive, an accelerated electron along a particular trajectory can be shown to emit classical electromagnetic radiation in the form of a Fermi-Dirac spectral distribution when observed in a particular angular regime. We investigate the relationship between the distribution, spectrum, and particle count. The result for the moving point charge is classical, as it accelerates along an exactly known trajectory.  We map to the semi-classical regime of the moving mirror model with a quantized spin-0 field. The scalars also possess a $\beta$ Bogoliubov coefficient distribution with Fermi-Dirac form in the respective frequency regime.

\end{abstract} 

\keywords{moving mirrors, black hole evaporation, acceleration radiation, Fermi-Dirac statistics}
\pacs{41.60.-m (Radiation by moving charges), 04.70.Dy (Quantum aspects of black holes)}
\date{\today} 

\maketitle


 \section{Introduction}
 It is well-known that bosons obey Bose-Einstein statistics, and fermions obey Fermi-Dirac statistics.  
Interestingly, Haro and Elizalde \cite{Haro:2008zza} found a result for the $\beta$-Bogoluybov coefficient of a semitransparent mirror demonstrating  
a flux of
scalar particles obeying a Fermi-Dirac distribution form in the large $\omega'$ limit.  Nicolaevici \cite{Nicolaevici:2009zz} confirmed the Fermi-Dirac form with respect to the energy $\omega$  but made special note that this does not establish the number of particles since it only applies in the large $\omega'$ limit. In a follow-up, Elizalde and Haro \cite{Elizalde:2010zza} recommended further investigation into the Fermi-Dirac form and its relationship to the sign 
in the $\beta$-Bogoliubov coefficient; in particular, the connection with the number of particles emitted per mode.

Here we investigate the situation using an ordinary moving point charge in classical electrodynamics \cite{Jackson:490457}.  We demonstrate the phenomenon without appealing to quantum field theory; i.e., one does not need to use moving mirrors or semi-transparency to understand the situation.  Nevertheless, we find the perfectly reflecting accelerating boundary corresponding to the moving point charge and examine its spectral statistics for clarity.  

The functional mapping between moving mirrors and moving point charges \cite{Ford:1982ct,Unruh:1982ic,Ritus:2003wu,Ritus:2002rq,Ritus:1999eu,Nikishov:1995qs,Zhakenuly:2021pfm,Ritus:2022bph} is leveraged to understand the problem in both contexts.  The situation has different physical meanings when examined for the classical electromagnetic field or quantized scalar field \cite{Ievlev:2023inj,Ievlev:2023bzk}; and thus, different implications for the classical radiation in ordinary 3+1 dimensions and quantum radiation in 1+1 dimensions. 

We use natural units, setting $\hbar = c = k_B = \mu_0 = 1$; the electron's charge is then a dimensionless number $e^2=4 \pi \alpha_{\textrm{fs}} \approx 0.092$. 

\section{Fermi-Dirac Trajectory}
\label{sec:FD}

\subsection{Dynamics and total energy}
\begin{figure}[h]
  \includegraphics[width=0.95\linewidth]{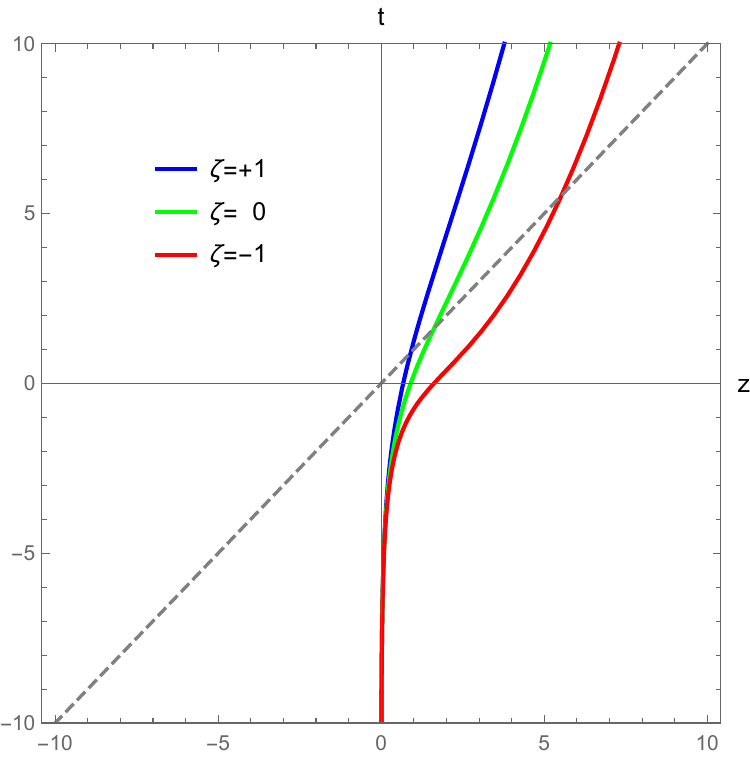}
 \caption{The position $z(t)$ trajectories, plotted to demonstrate various $\zeta$ values of Eq.~(\ref{eom}).  The key takeaway is the motions are asymptotic resting and are limited to the half-plane $z>0$.  }
\label{FDspacetime}
\end{figure}

\begin{figure}[h]
  \includegraphics[width=0.95\linewidth]{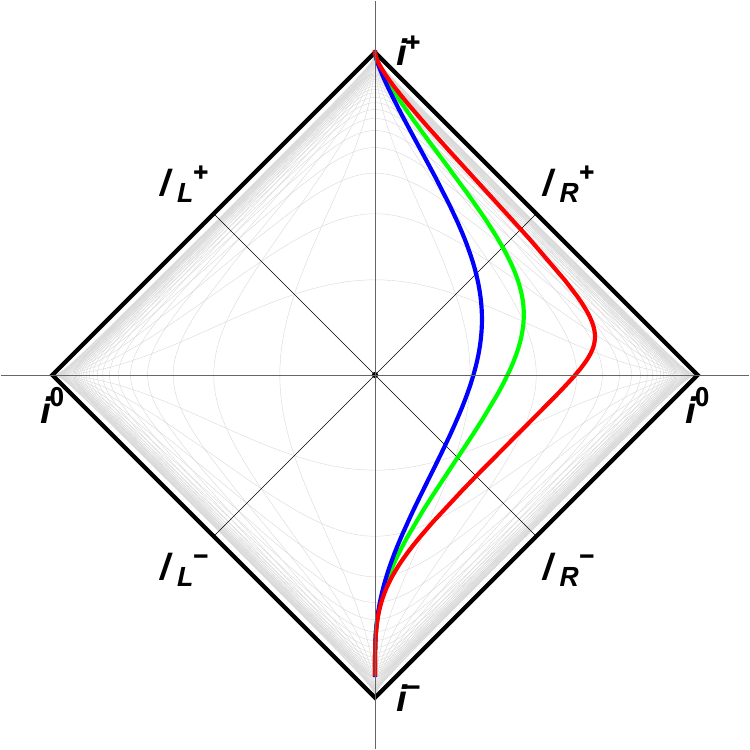}
 \caption{The position $z(t)$ trajectories, plotted in a Penrose diagram to demonstrate the $\zeta = (-1,0,+1)$ values of Eq.~(\ref{eom}).  The key takeaway is the motions are asymptotic resting and are limited to the half-plane $z>0$. The same color scheme of Fig. \ref{FDspacetime} is used; here, $\kappa =2$ for illustration. }
\label{FDpenrose}
\end{figure}

Let us start with a simple illustration of the situation.  Consider an electron moving in a straight line along the $z$-axis.
We take the trajectory defined implicitly as 
\be t(z) = \frac{\kappa}{4} z^2 + \frac{2}{\kappa} \ln ( \kappa z ) + z \zeta,\label{eom}\ee
where $\kappa > 0$ is  the acceleration scale and $-1 < \zeta < 1$.
The inverse velocity along the trajectory and the maximum velocity are, respectively,
\begin{equation}
    \frac{1}{v} = \frac{\diff t(z)}{\diff z} = \frac{\kappa  z}{2}+\frac{2}{\kappa  z} + \zeta \,, \quad
    v_\text{max} = \frac{1}{2 + \zeta} \,.
\label{velocity}
\end{equation}
From Eq.~\eqref{velocity}, it is evident that for $\zeta > -1$ this trajectory travels along a time-like, relativistic worldline.
This trajectory is asymptotically static; see Fig. \ref{FDspacetime} for a spacetime diagram and Fig. \ref{FDpenrose} for a Penrose diagram.

\begin{figure}[h]
  \includegraphics[width=0.95\linewidth]{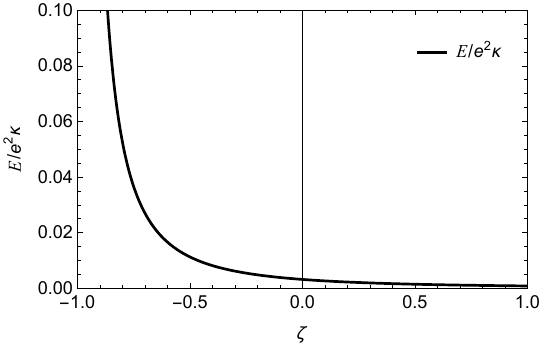}
 \caption{Total radiated energy $E / e^2 \kappa$ as a function of parameter $\zeta$. The energy blows up as $\zeta \to -1$; as this corresponds to the speed of light for the trajectory's maximum velocity, e.g. see the red worldline in Fig. \ref{FDspacetime} and Fig. \ref{FDpenrose}. }
\label{fig:TotalEnergy}
\end{figure}

The total energy emitted can be calculated with the Larmor formula; this energy is finite for $\zeta > -1$.
For example, when $\zeta = 0$ it takes the analytic form:
\be E = \frac{e^2\kappa}{36}\left(\frac{1}{3\sqrt{3}} - \frac{1}{4\pi}\right).\label{totenergy}\ee
For other values of the parameter $\zeta$, the analytic formula for the total energy exists but is complicated; nevertheless, it is simple to illustrate numerically, see Fig.~\ref{fig:TotalEnergy}.
One can use the total energy, e.g. Eq.~(\ref{totenergy}), to check the consistency of the spectral results (see the Appendix for detail).  

\subsection{Spectral distribution of the accelerating electron's radiation}
To find the spectral distribution for Eq.~(\ref{eom}), we use the standard approach in classical electrodynamics \cite{Jackson:490457}. Here the energy $E$ can be found by the spectrum $I(\omega)$, or spectral distribution $\diff{I}/\diff{\Omega}$ 
 \cite{Schwinger1949},
\be E = \int \diff{\omega} I(\omega) = \int \diff{\omega} \int \diff{\Omega} \frac{\diff{I(\omega)}}{\diff{\Omega}}.\ee
For example, by the use of Eq. 13 in \cite{Ievlev:2023inj}, 
\begin{equation}
	\frac{\diff I(\omega)}{\diff \Omega} = \frac{e^2\omega^2}{16\pi^3}\left|\sin\theta \int\displaylimits_{0}^{\infty} \diff z  e^{i \phi(z)}  \right|^2,
\label{I_phi_int}
\end{equation}
where $\phi = \omega (t - z\cos\theta)$, we write
\begin{equation}
	\frac{1}{\sin^2\theta}\frac{\diff I(\omega)}{\diff \Omega} = \frac{e^2\omega^2}{16\pi^3}\left| \int\displaylimits_{0}^{\infty} \diff z  e^{i \phi(z)}  \right|^2.
\label{I_phi_int1}
\end{equation}
This integral can be solved exactly as is (see the Appendix), but for simplicity, consider the spectral distribution at a particular angle $\theta_0$ instead.  Specialize to $\theta \to \theta_0$ such that the phase is 
\be \phi(z) = \frac{\kappa \omega}{4}z^2 + \frac{2}{\kappa} \omega \ln z + \omega z (\zeta  - \cos\theta_0).\label{phase}\ee
It is now straightforward to integrate Eq.~(\ref{I_phi_int1}) when $\cos \theta_0 = \zeta$.  One obtains the spectral distribution,
\begin{equation}
	\frac{1}{\sin^2\theta_0}  \left.\frac{\diff I(\omega)}{\diff \Omega}\right|_{\theta_0} = \frac{e^2 }{8\pi^2}\frac{\omega/\kappa}{e^{2\pi \omega/\kappa}+1},
\label{I_phi_int2}
\end{equation}
where the spectral frequency content of the radiation has a Fermi-Dirac form. Before examining this form and its relationship to its particle spectrum $N(\omega)$, let us first look at its quantum dual in the moving mirror model \cite{DeWitt:1975ys,Davies:1976hi,Davies:1977yv}, in the spirit of previous moving mirror studies on the Fermi-Dirac result \cite{Haro:2008zza,Nicolaevici:2009zz,Elizalde:2010zza}.

\subsection{Corresponding Bogolubov Coefficients }

While the above result is classical radiation from the 3+1 dimensional electromagnetic field, we can investigate the quantum radiation from the 1+1 dimensional scalar field of the moving mirror model. 
The mapping recipe \cite{Ievlev:2023inj} between electron and mirror links the spectral distribution on the electron side and the Bogolubov coefficient squared on the mirror side, c.f. \cite{Ievlev:2023bzk}:
\begin{equation}
\begin{aligned}
	&\frac{\diff{I}}{\diff{\Omega}}(\omega,\cos\theta) = \frac{e^2 \omega^2}{4\pi} |\beta_{pq}|^2, \\
	&p + q = \omega \,, \quad p - q = \omega \cos\theta. \\
\end{aligned}
\label{recipe_dIdOmega_from_mirror}
\end{equation}
Using the full spectral distribution from Eq.~\eqref{exact} one can use this recipe to obtain the corresponding Bogolubov coefficients.
Let us however consider setting $\theta = \theta_0$, $\zeta =\cos\theta_0$.
In this case, the electron's spectral distribution has a simple form Eq.~\eqref{I_phi_int2}.
In terms of the scalar frequencies $p,q$, the corresponding condition reads
\begin{equation}
    \theta = \theta_0 \,, \ \zeta =\cos\theta_0
    \Longleftrightarrow
    p = \omega \frac{1 + \zeta}{2} \,, \ q = \omega \frac{1 - \zeta}{2}
\label{theta_condition_pq_correspondence}
\end{equation}
So, the scalar frequencies $p$ and $q$ are not independent, but related to each other through this condition.
The recipe Eq.~\eqref{recipe_dIdOmega_from_mirror} gives particular beta Bogolubov coefficients:
\begin{equation}
    |\beta_{pq}|^2 = \frac{1-\zeta^2}{2\pi (p+q) \kappa}\frac{1}{e^{2\pi  (p+q)/\kappa }+1} \,, \quad
    \frac{p}{q} \equiv \frac{1 + \zeta}{1 - \zeta} \,.
\label{pandq}
\end{equation}
%
%
This result demonstrates the Fermi-Dirac form for the $\beta$-Bogolubov coefficients of the quantum scalars.  

By tuning the value of $\zeta$ one can obtain a trajectory that gives the Fermi-Dirac at any pre-assigned angle (or a bespoke frequency regime).
For instance, 
the `high-frequency' regime \cite{Hawking:1974sw}, $p \sim 0$ ($q\gg p$), corresponds to $\zeta \sim -1$; 
Eq.~(\ref{pandq}) becomes to leading order
\be |\beta_{pq}|^2 = \frac{1+\zeta}{\pi q \kappa}\frac{1}{e^{2\pi  q/\kappa }+1} \,. \label{betaFD}\ee
Using the duality to map back to the electron, the choice $\zeta = \cos\theta_0 = -1$ corresponds to a viewpoint behind the accelerating electron $\theta_0 \sim \pi$. 

\subsection{Connection to Particle Count}
The notion of discrete radiation energy $\hbar \omega$ 
allows an introduction of a 
particle spectrum $N(\omega)$. The connection between the spectral distribution of electromagnetic waves and the particle spectrum is \cite{Jackson:490457}: 
\be N(\omega) = \frac{1}{\omega} I(\omega) = \frac{1}{\omega} \int d\Omega \frac{\diff{I}}{\diff{\Omega}},\label{classicalparticles}\ee
which must be consistent with the total energy emission as computed by the spectral distribution,
\be E = \int \diff{\omega} \; \omega N(\omega) = \int \diff{\omega} \int \diff{\Omega} \frac{\diff{I}}{\diff{\Omega}}.\ee
Therefore, the Fermi-Dirac distribution does not correspond to the particle spectrum $N(\omega)$, or the energy spectrum $I(\omega)$. 
It does not even correspond to the spectral distribution $\diff{I}/{\diff{\Omega}}$ at an arbitrary observation angle $\theta$; 
but only in a specific angular regime $\theta \to \theta_0$: $\left.\diff{I}/{\diff{\Omega}}\right|_{\theta_0}$ using the corresponding trajectory, Eq.~(\ref{eom}).  Nevertheless, the interesting question remains if it is possible to observe such radiation measured in such a specified angular regime $\theta_0$.  Does an observer see Fermionic electromagnetic radiation? Is the spectral content congruent with Fermi-Dirac statistics?

We stress again that the trajectory is easily generalized in a number of different ways to illustrate the robust Fermi-Dirac form of the spectral distribution at particular angles.  For instance, a particular choice of $\zeta$ in Eq.~(\ref{eom}) results in a new trajectory form, capable of a new observation angle $\theta_0$, which gives the Fermi-Dirac result. This means, depending on the particular bespoke trajectory of interest, the relevant zeta-angle $(\zeta,\theta_0)$ could be in any desired direction, such as to the side $ (0,\pi/2)$, in front $(+1,0)$, or behind $(-1,\pi)$ the accelerating electron.

\section{Discussion}
The particular physics of this result depends in subtle ways on dimension (3+1 vs 1+1), source (electron vs. mirror), and regime (angle vs. frequency). 
Two other notable examples in the literature confirm this, namely, scalar charges and even-odd dimensional dependence. 
Let us consider these examples now.

For an example of a source other than an electric charge or mirror, Nikishov and Ritus found \cite{Nikishov:1995qs} that in the case of a scalar charge, the emitted scalar radiation along a particular trajectory will obey Fermi-Dirac statistics; in contrast to an electric charge following the same particular trajectory whose spin-1 radiation field obeys Bose-Einstein statistics.  

As an example of dimensional dependence, scalar field radiation measured by a uniformly accelerated DeWitt detector obeys Bose-Einstein statistics when the dimension of the spacetime is even, but when the dimension is odd, one obtains Fermi-Dirac statistics \cite{Takagi:1986kn}.

Taken as a whole, it is clear the subtleties involved make it especially important to precisely define the context and the regime of applicability and explicitly examine the form of the computed observables. 

We note that Hawking radiation \cite{Hawking:1974sw} and its Schwarzschild moving mirror analog \cite{Good:2016oey} utilize the high-frequency regime lending support to the notion of thermality and Bose-Einstein distributed scalars at late-times. 
However, in the Schwarzschild case (as opposed to extremal cases \cite{good2020extreme} or asymptotically inertial situations \cite{Good:2019tnf}), there is no finite total energy check corresponding to the Bogoliubov coefficients. Moreover, the total particle count is infinite. This is ultimately due in part to the horizon; and in the analog moving mirror situation, the fact that the proper acceleration is asymptotically infinite. The same goes for the eternal black hole analog of Carlitz-Willey \cite{carlitz1987reflections}. 

In this work, we have shown that the scalars can possess Fermi-Dirac distributed $\beta$-Bogliubov coefficients.  If one considers $\beta$-Bogoliubov coefficients sufficient evidence of a thermal Bose-Einstein distribution for the Schwarzschild or Carlitz-Willey trajectories; then the result Eq.~(\ref{betaFD}) is also sufficient evidence of a thermal Fermi-Dirac distribution for the trajectories Eq.~(\ref{eom}). Moreover, we have an additional check of total finite energy and finite particle emission (see the Appendix for more detail).   

This result demonstrates that, although the high-frequency approximation (or the low-frequency approximation, if one prefers) is frequently used in the literature, one should clearly understand whether this approximation represents the physical system under consideration.
The result Eq.~\eqref{betaFD} does not reveal e.g. the particle spectrum, as the high-frequency region does not dominate the corresponding contribution from the beta Bogolubov coefficients.
In other words, one should be careful when applying the high-frequency (or the low-frequency) approximation; in each case, this approximation should be well-motivated; otherwise, peculiar results may arise.

\section{Conclusion}
We have shown that moving point charge radiation can possess a Fermi-Dirac spectral distribution form.  A particular trajectory and corresponding angular regime demonstrate the result.  By appealing to classical electrodynamics, we have analyzed the physical reason for the resulting unexpected spectral-statistics form.  

The spectral-statistics (as explicitly derived from the spectral distribution in a particular angular regime for the radiation from a moving point charge) do not necessarily characterize the spin-statistics of the electromagnetic field in question.  Instead, they depend crucially on the observation angle and the specific electron trajectory interaction with the radiation field.

\section{Acknowledgements} 
Funding comes in part from the FY2021-SGP-1-STMM Faculty Development Competitive Research Grant No. 021220FD3951 at Nazarbayev University.   

\appendix

\section{Partial contribution from the FD particles}

The partial energy contribution when $\zeta = \cos\theta_0$ can be found from the Fermi-Dirac form Eq.~(\ref{I_phi_int2}) of the electron,
\begin{equation}
    E^{fd}_\textrm{electron} 
        = \int_0^\infty \diff{\omega} \int_0^{2\pi} \diff{\varphi} \left.\frac{\diff I(\omega)}{\diff \Omega}\right|_{\theta_0} 
        = \frac{e^2\kappa (1 - \zeta^2) }{192\pi}
\label{partial_energy_electron}
\end{equation}
The corresponding contribution from the mirror can be derived from this result in the following way. 
We insert into Eq.~\eqref{partial_energy_electron} the integral
$1 = \int_{-1}^{1} \diff (\cos\theta) \, \delta( \cos\theta - \zeta ) $ 
and change the variables according to Eq.~\eqref{recipe_dIdOmega_from_mirror}. The Jacobian is 
$2 / (p + q)$, and the resulting contribution is
\begin{equation}
\begin{aligned}
    &E^{fd}_\textrm{mirror}  \\
        &= \int_0^\infty \diff{p}\int_0^\infty \diff{q}\; 
            \delta\left( \frac{p-q}{p+q} - \zeta \right) \,
            (p + q) |\beta_{pq}|^2 \\
        &= \frac{\kappa (1 - \zeta^2) }{192\pi}.
\end{aligned}
\end{equation}
This gives the partial contribution to the energy emitted to both sides of the corresponding mirror, counting only the FD particles.

Let us look at how the analogy extends to finite particle count.  
For the moving mirror, the total number of scalars emitted to the right side of the mirror is:
\begin{equation}
\begin{aligned}
    &N_\textrm{mirror} \\
    &= \int_0^\infty \diff{p}\int_0^\infty \diff{q}\; 
        \delta\left( \frac{p-q}{p+q} - \zeta \right) \,
        |\beta_{pq}|^2 \\
    &= \frac{(1 - \zeta^2) \ln 2 }{8\pi^2} \,.
\end{aligned}
\label{bo}
\end{equation}
%
%
For the case of the accelerating electron, we may integrate over all frequencies $\omega$ on Eq.~(\ref{classicalparticles}) which gives
\begin{equation}
\begin{aligned}
    \int_0^\infty \diff{\omega}\; N(\omega) 
        &= \int_0^\infty \frac{\diff{\omega}}{\omega} \int_0^{2\pi} \diff{\psi} \left.\frac{\diff I(\omega)}{\diff \Omega}\right|_{\theta_0},\\
        &= e^2\frac{(1 - \zeta^2) \ln 2 }{8\pi^2},
\end{aligned}
\end{equation}
in agreement with Eq.~(\ref{bo}). This highlights the dual consistency between the mirror and electron but also demonstrates the advantage of a finite energy and finite particle count with an exact analytic solution, Eq.~(\ref{eom}).


\section{Exact spectrum}

To demonstrate the difference between a particular choice of observation angle and the spectrum $I(\omega)$ which results from integration over the solid angle, we briefly look at an exact answer for the integral of Eq.~(\ref{I_phi_int1}), setting $\zeta = 0$ and leaving $\theta$ unset.  Integrating Eq.~(\ref{I_phi_int1}) gives
\begin{equation}
\begin{aligned}
	&\frac{\diff I(\omega)}{\diff \Omega} =\frac{e^2  \omega \sin^2\theta }{16 \pi ^3 \kappa}  e^{-\frac{\pi  \omega }{\kappa }} \times \\
	&\times \left| \Gamma \left(\frac{1}{2}-\frac{i \omega }{\kappa }\right)A+2 \cos \theta \sqrt{\frac{i \omega}{\kappa}} \Gamma \left(1-\frac{i \omega }{\kappa }\right)B\right|^2 \,,
\end{aligned}
\label{exact}
\end{equation}
where 
\begin{equation}
\begin{aligned}
	A &= \, _1F_1\left(\frac{1}{2}-\frac{i \omega }{\kappa };\frac{1}{2};\frac{i \omega \cos^2\theta }{\kappa }\right) \,, \\
	B &= \, _1F_1\left(1-\frac{i \omega }{\kappa };\frac{3}{2};\frac{i\omega \cos^2\theta  }{\kappa }\right) \,.
\end{aligned}
\end{equation}
This spectral distribution, Eq.~(\ref{exact}), gives the total energy, Eq.~(\ref{totenergy}) by numerical integration,
\begin{equation}
\begin{aligned}
	E &= \int_0^{\infty} \diff{\omega} \int_0^{2\pi}\diff{\phi}\int_0^\pi \diff{\theta}\sin\theta \frac{\diff I(\omega)}{\diff \Omega} \\
	&= \frac{e^2\kappa}{36}\left(\frac{1}{3\sqrt{3}} - \frac{1}{4\pi}\right) \,.
\end{aligned}
\end{equation}
We cannot integrate Eq.~(\ref{exact}) exactly over the solid angle to obtain an analytic form of $I(\omega)$. However, it is clear that Eq.~(\ref{exact}) will not result in a Fermi-Dirac form for the spectrum $I(\omega)$, even though at $\theta_0 =\pi/2$, Eq.~(\ref{exact}) gives 
\begin{equation}
	\left.\frac{\diff I(\omega)}{\diff \Omega}\right|_{\theta_0} = \frac{e^2 }{8\pi^2}\frac{\omega/\kappa}{e^{2\pi \omega/\kappa}+1},
\end{equation}
which is the result Eq.~(\ref{I_phi_int2}).  For this reason, one cannot say the particle count, $N(\omega)$ rests in a Fermi-Dirac distribution, which is ultimately consistent with the horizonless globally defined motion, Eq.~(\ref{eom}), evolving to an asymptotic stop.

\bibliography{main} 
\end{document}